# NP-Hardness of Approximating Nash Social Welfare with Supermodular Valuations

Alon Bebchuk, Tel Aviv University

**ABSTRACT**

We study the problem of allocating a set of indivisible items to agents with supermodular utilities to maximize the Nash social welfare. We show that the problem is NP-hard for any approximation factor.

## 1 INTRODUCTION

The Nash social welfare (NSW) problem is that of allocating a set $B$ of indivisible items among a set $A$ of agents, $alloc = \{B_a\}_{a \in A}$, where each agent $a \in A$ has a valuation function $v_a: 2^B \to R_{\geq 0}$, such that the geometric mean of agent valuations for the allocation, $NSW(alloc) = (\prod_{a \in A} v_a(B_a))^{\frac{1}{|A|}}$, is maximal. For constant $c \geq 1$, allocation $alloc$ is said to be a $c$-approximate solution to the NSW problem if $NSW(alloc) \geq \frac{OPT}{c}$, where $OPT$ is the optimum value of the NSW-maximization problem.

The problem of allocating resources in a manner which is both efficient and fair is a fundamental problem spanning the fields of economics, social choice theory, and computer science with a significant amount of literature existing for each [2, 5, 6, 20, 21, 22, 23]. While Utilitarian social welfare (USW), i.e. the problem of maximizing the sum of agent valuations, is a common measure of efficiency, it does not achieve any guarantee of fairness, as all resources may be allocated to a single agent if his valuation for them is high enough. On the other hand, Egalitarian social welfare, i.e. the problem of maximizing the minimum over agent valuations, which is a measure of fairness, is pareto-inefficient by Harsanyi's utilitarian theorem [16]. As shown in [7], NSW over indivisible goods strikes a sweet balance between efficiency and fairness. NSW has also been shown to possess several desirable features [20], including scale-freeness, as scaling an agent's valuation function does not affect the outcome.

## 2 PRELIMINARIES

The complexity of the Nash social welfare problem depends on the class of agents' valuation functions. In this paper, we will focus on the setting where agents' valuation functions are supermodular. We also require that valuation functions are normalized, i.e. $v(\emptyset) = 0$, and monotone, i.e. $S \subseteq T \Rightarrow v(S) \leq v(T)$.

*Classes of valuation functions.*

1. A valuation $v$ is additive if $v(S) = \sum_{j \in S} v(\{j\})$ for any $S \subseteq B$.
2. A valuation $v$ supermodular if $v(S \cup T) \geq v(S) + v(T) - v(S \cap T)$ for any $S, T \subseteq B$.
3. A valuation $v$ is superadditive if $v(S \cup T) \geq v(S) + v(T)$ for any $S, T \subseteq B$ such that $S$ and $T$ are disjoint.

The above classes of valuation functions form a chain of inclusions: additive valuations are supermodular and supermodular valuations are superadditive.

## 3 PRIOR WORK

Through a reduction from Subset-Sum the Nash social welfare problem is known to be NP-hard even in the case where there are two agents with identical additive valuations. Lee [17] showed that the Nash social welfare problem for additive valuations is NP-hard to approximate withing a factor of 1.00008. Shortly after, Garg, Hoefer, and Mehlhorn [12] improved this lower bound to $\sqrt{\frac{8}{7}} > 1.069$. It is generally believed approximating the NSW is than the Utilitarian social welfare, however, up until now no formal reduction has been made.

The first constant-factor approximation algorithm for additive valuations, given by Cole and Gkatzelis [9], provides an approximation factor of $2e^{\frac{1}{e}} \approx 2.889$. Afterwards [8], an improved analysis of this algorithm improved this factor to 2. As of now the best approximation factor known is $e^{\frac{1}{e}} \approx 1.445$ by Barman, Krishnamurthy, and Vaish [9]. This result give a clear separation between the hardness of the additive and supermodular settings, with the later shown to be NP-hard to approximate within a factor of $\frac{e}{e-1} > 1.581$ [13].

Further complementing the results for additive functions, constant-factor approximations were also extended to the following classes: capped-additive [12], SPLC [1], and a common generalization of both, capped-SPLC [8], with the approximation factor for capped-additive matching the best known $e^{\frac{1}{e}} \approx 1.445$ factor for additive valuations. Additionally, an algorithm for approximating the optimal NSW value, without finding a solution, with a factor of $\frac{(e-1)^2}{e^3}$ for a broad subclass of submodular functions was given by Li and Vondrak [19].

A constant-factor approximation algorithm for "Rado valuations" [14] provided a breakthrough which was followed by the first constant-factor approximation algorithm for supermodular-valuations [18], with an approximation factor of 380. This factor was later improved to $4 + \epsilon$ [13].

For the more general class of XOS and even more general class of subadditive valuations, up until recently only polynomial approximation factors were achieved [3]. In a recent paper [11], Dobzinski, Li, Rubinstein and Vondrak presented a first constant-approximation algorithm in the demand query model for subadditive valuations. The factor itself is quite large, and further research is required to close the gap between the lower bound on approximation and achieved approximation results.

Prior to this paper no results regarding the complexity of supermodular and superadditive valuations has appeared in the literature.

## 4  MAIN RESULT

Our main result is the following.

**Theorem.** It is NP-hard to approximate the Nash social welfare with supermodular valuations for any approximation factor.

To achieve this result, we present a polynomial time reduction from the problem of Vertex Cover on 3-regual graphs, a well-known NP-complete problem [15], to that of $c$-approximating the NSW with supermodular valuations, for any constant $c \geq 1$. The reduction draws inspiration from the reduction presented by Lee [17] in proving that maximizing NSW with additive valuations is APX-hard.

## 5 REDUCTION

In this section, we design the reduction from the Vertex Cover problem over 3-regular graphs to the problem of $c$-approximating the maximal Nash social welfare in a combinatorial auction where all but one agent have additive valuations, and the remaining agent has a supermodular valuation, for any $c \geq 1$. Proof of correctness and implications will be provided in the subsequent sections.

Given a 3-regular graph $G = (V, E)$ and a constant $c \geq 1$, the reduction proceeds as follows:

- **Number the edges**: Assign numbers to the edges in $E$.
- **Define items for each vertex**: For each vertex $v \in V$, create three distinct items $\{v', v'', v'''\}$.
- **Denote incident edges for each vertex**: Let $E_v$ denote the set of edges incident to vertex $v$. Since $G$ is 3-regular, each vertex has exactly 3 incident edges. Denote these edges as $e'_v$ (the incident edge with the smallest number), $e''_v$ (the incident edge with the middle number), and $e'''_v$ (the incident edge with the largest number).
- **Map edges to vertex items**: For each vertex $v \in V$, define a one-to-one mapping $f_v$ that assigns the edges incident to $v$ to the items $\{v', v'', v'''\}$. Specifically, $f_v$ maps $e'_v$ to $v'$, $e''_v$ to $v''$, and $e'''_v$ to $v'''$.
- **Define edge agents**: For each edge $e \in E$, define an edge agent $a_e$. The valuation function $v_e$ of agent $a_e$ is an additive valuation function, where $v_e(\{j\}) = \begin{cases} 1, & j \in \{f_v(e) | v \in e\} \\ 0, & otherwise \end{cases}$. In words, the utility of agent $a_e$ for a bundle $S$ is the number of vertex items mapped to edge $e$ that are contained in $S$.
- **Define the greedy agent**: Define the greedy agent $g$. Denote $V_S$ as the set of vertices whose items are all contained in bundle $S$, i.e. $V_S = \{v \in V | v', v'', v''' \in S\}$. These vertices are referred to as the vertices covered by S. Set a constant $\alpha = 8^N \cdot (c + \epsilon)^{\frac{3}{2}N+1}$, where $\epsilon$ is an arbitrarily small positive value. The reasoning behind this choice of $\alpha$ will be clarified in the proofs. The valuation function of agent $g$ is defined as, $v_g(S) = \alpha^{|V_S|}$.

We will show that $v_g$ is supermodular. The following table shows the relationship between $\mathbb{1}_{v \in V_{(S \cap T)}}$ and $\mathbb{1}_{v \in V_{(S \cup T)}}$, relative to $\mathbb{1}_{v \in V_S}$ and $\mathbb{1}_{v \in V_T}$.

| $\mathbb{1}_{v \in V_S}$ | $\mathbb{1}_{v \in V_T}$ | $\mathbb{1}_{v \in V_{(S \cap T)}}$ | $\mathbb{1}_{v \in V_{(S \cup T)}}$ |
| --- | --- | --- | --- |

| 0 | 0 | 0 | 0 or 1 |
| 0 | 1 | 0 | 1 |
| 1 | 0 | 0 | 1 |
| 1 | 1 | 1 | 1 |

**Table 5.1**: In the top-right entry, $\mathbb{1}_{v \in V_{(S \cup T)}}$ is 0 or 1 depending on whether the union $S \cup T$ includes all three items corresponding to vertex $v$. For example, if $S = \{v'\}$ and $T = \{v''\}$, then $S \cup T = \{v', v''\}$, and $v$ is not covered by $S \cup T$, resulting in 0. However, if $S = \{v'\}$ and $T = \{v'', v'''\}$, then $S \cup T = \{v', v'', v'''\}$, and $v$ is covered, resulting in 1.

From the table, we observe that $\mathbb{1}_{v \in V_{(S \cup T)}} \geq \mathbb{1}_{v \in V_S} + \mathbb{1}_{v \in V_T} - \mathbb{1}_{v \in V_{(S \cap T)}} \geq 0$. Since $\alpha \geq 1$, we have,

$$\alpha^{\sum_{v \in V} \mathbb{1}_{v \in V_{(S \cup T)}}} \geq \alpha^{\sum_{v \in V} \mathbb{1}_{v \in V_S} + \sum_{v \in V} \mathbb{1}_{v \in V_T} - \sum_{v \in V} \mathbb{1}_{v \in V_{(S \cap T)}}} \geq \alpha^{\sum_{v \in V} \mathbb{1}_{v \in V_S} + \sum_{v \in V} \mathbb{1}_{v \in V_T}}.$$

Let $a = \alpha^{\sum_{v \in V} \mathbb{1}_{v \in V_S}}$ and $b = \alpha^{\sum_{v \in V} \mathbb{1}_{v \in V_T}}$, where $a, b \geq 1$. Therefore, the inequality $(a - 1) \cdot (b - 1) \geq 0$ holds. Expanding this inequality gives $ab \geq a + b - 1$. Thus, we conclude that,

$$\alpha^{\sum_{v \in V} \mathbb{1}_{v \in V_S} + \sum_{v \in V} \mathbb{1}_{v \in V_T}} \geq \alpha^{\sum_{v \in V} \mathbb{1}_{v \in V_S}} + \alpha^{\sum_{v \in V} \mathbb{1}_{v \in V_T}} - 1.$$

Since

$$\alpha^{\sum_{v \in V} \mathbb{1}_{v \in V_S}} + \alpha^{\sum_{v \in V} \mathbb{1}_{v \in V_T}} - 1 \geq \alpha^{\sum_{v \in V} \mathbb{1}_{v \in V_S}} + \alpha^{\sum_{v \in V} \mathbb{1}_{v \in V_T}} - \alpha^{\sum_{v \in V} \mathbb{1}_{v \in V_{(S \cap T)}}}$$

we can deduce that

$$\alpha^{\sum_{v \in V} \mathbb{1}_{v \in V_{(S \cup T)}}} \geq \alpha^{\sum_{v \in V} \mathbb{1}_{v \in V_S}} + \alpha^{\sum_{v \in V} \mathbb{1}_{v \in V_T}} - \alpha^{\sum_{v \in V} \mathbb{1}_{v \in V_{(S \cap T)}}}.$$

From the definition, $v_g(S) = \alpha^{\sum_{v \in V} \mathbb{1}_{v \in V_S}}$. Therefore, for any $S, T \subseteq B$,

$$v_g(S \cup T) = \alpha^{\sum_{v \in V} \mathbb{1}_{v \in V_{(S \cup T)}}},$$

and

$$v_g(S) + v_g(T) - v_g(S \cap T) = \alpha^{\sum_{v \in V} \mathbb{1}_{v \in V_S}} + \alpha^{\sum_{v \in V} \mathbb{1}_{v \in V_T}} - \alpha^{\sum_{v \in V} \mathbb{1}_{v \in V_{(S \cap T)}}}.$$

Thus, $v_g(S \cup T) \geq v_g(S) + v_g(T) - v_g(S \cap T)$, proving that $v_g$ is supermodular.

This reduction results in a combinatorial auction with $n = \frac{3}{2}N + 1$ agents (comprising $\frac{3}{2}N$ additive edge agents and 1 supermodular greedy agent) and $m = 3N$ items. The reduction is computable in polynomial time.

**Example 5.2.**

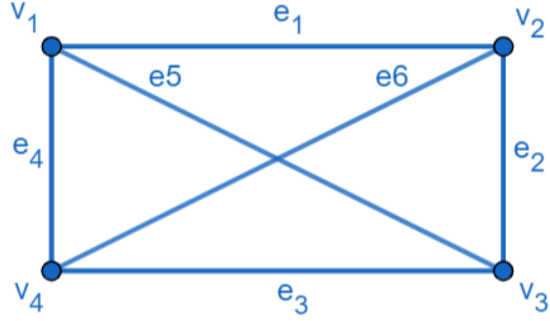

Consider the 3-regular graph $G = (\{v_1, v_2, v_3, v_4\}, \{e_1, e_2, e_3, e_4, e_5, e_6\})$. The reduction produces the following combinatorial auction:

- The set of agents is $A = \{a_{e_1}, a_{e_2}, a_{e_3}, a_{e_4}, a_{e_5}, a_{e_6}, g\}$.
- The set of items is $B = \{v_1', v_1'', v_1''', v_2', v_2'', v_2''', v_3', v_3'', v_3''', v_4', v_4'', v_4'''\}$.
- The valuation functions for the edge agents are
$$v_{e_1}(S) = |S \cap \{v_1', v_2'\}|, \quad v_{e_2}(S) = |S \cap \{v_2'', v_3'\}|, \quad v_{e_3}(S) = |S \cap \{v_3'', v_4'\}|,$$
$$v_{e_4}(S) = |S \cap \{v_1'', v_4''\}|, \quad v_{e_5}(S) = |S \cap \{v_1''', v_3'''\}|,$$
$$v_{e_6}(S) = |S \cap \{v_2''', v_4'''\}|.$$
- The greedy agent's valuation function is $v_g(S) = \alpha^{|\{v \in V | v', v'', v''' \in S\}|}$.

**Notation 5.3.** For any allocation $alloc = \{B_e\}_{e \in E} \cup \{B_g\}$, let $V_E$ denote the set of vertices such that at least one of their vertex items is allocated to an edge agent, i.e. $V_E = \{v \in V | \{v', v'', v'''\} \cap (\cup_{e \in E} B_e) \neq \emptyset\}$. Similarly, let $V_g$ denote the set of vertices covered by $B_g$. From these definitions, we have,

$$V_E = \{v \in V | \{v', v'', v'''\} \cap (\cup_{e \in E} B_e) \neq \emptyset\} = V / \{v \in V | v', v'', v''' \in B_g\} = V / V_g.$$

Therefore, $V_E$ and $V_g$ are complementary.

The key insight from the reduction is the tradeoff between $V_E$ and $V_g$. For the Nash social welfare to be positive, each edge agent must receive at least one vertex item mapped to it, meaning that $V_E$ must form a vertex cover. Simultaneously, maximizing $V_g$ increases the utility of the greedy agent exponentially by a factor of $\alpha$, creating a balance between minimizing the vertex cover and maximizing the Nash social welfare.

## 6     MAIN THEOREM

**Theorem 6.1.** Given a 3-regular graph $G = (V, E)$ and a constant $c \geq 1$, let $alloc = \{B_e\}_{e \in E} \cup \{B_g\}$ be an allocation providing a $c$-approximation of the maximal Nash social welfare in the auction produced by the reduction described in Section 5. Then, $V_E$ is a minimum vertex cover of $G$.

**Corollary 6.2.** The problem of maximizing Nash social welfare in a combinatorial auction where all but one agent have additive valuations, and the remaining agent has a supermodular valuation is NP-hard for any approximation factor.

*Proof.*

From Theorem 6.1, there exists a polynomial-time reduction from the NP-complete Vertex Cover problem over 3-regular graphs to the problem of $c$-approximating maximal Nash social welfare in a combinatorial auction where all but one agent have additive valuations, and the remaining agent has a supermodular valuation. This implies that the reduced problem is NP-hard. Moreover, since the reduction holds for any $c \geq 1$, it follows that the reduced problem is NP-hard for any approximation factor. ∎

## 7 PROOF OF MAIN THEOREM

**Definition 7.1.** A vertex cover of a graph is a set of vertices that includes at least one endpoint of every edge of the graph; a minimal vertex cover is a vertex cover that is not a strict subset of any other vertex cover; and a minimum vertex cover is a vertex cover with the smallest possible number of vertices.

**Notation 7.2.** Given a vertex set $S \subseteq V$, denote $d_S$ as the number of edges where both endpoints are contained in $S$.

**Lemma 7.3.** Let $alloc = \{B_e\}_{e \in E} \cup \{B_g\}$ be an allocation with positive Nash social welfare. Then $V_E$ is a vertex cover of $G$, and the Nash social welfare of the allocation is equal to $\left(2^k \cdot \alpha^{N-|V_E|}\right)^{\frac{1}{n}}$, for some $0 \leq k \leq d_{V_E}$.

*Proof.*

From the assumption on $alloc$, for each edge $e \in E$, $|B_e \cap I_e| = v_e(B_e) \in \{1,2\}$. Denote the endpoints of $e$ by $v_i, v_j$. Then, by construction, $V_E \cap \{v_i, v_j\} \neq \emptyset$, meaning that at least one of the endpoints of each edge $e$ is contained in $V_E$, and thus, $V_E$ is a vertex cover of $G$.

From the definition, $|V_g| = |V/V_E| = N - |V_E|$, and therefore, $v_g(B_g) = \alpha^{N-|V_E|}$. If for edge $e \in E$, $v_e(B_e) = 2$, then both endpoints of edge $e$ are contained in $V_E$. Therefore, the number of edge agents with utility of 2, denoted by $k$, is at most $d_{V_E}$. Altogether, the Nash social welfare of the allocation is $NSW(alloc) = \left(2^k \cdot \alpha^{N-|V_E|}\right)^{\frac{1}{n}}$. ∎

**Lemma 7.4.** Let $alloc = \{B_e\}_{e \in E} \cup \{B_g\}$ be an allocation with positive Nash social welfare such that $V_E$ is not a minimal vertex cover of $G$. Then $alloc$ does not provide a $c$-approximation of the maximal Nash social welfare.

*Proof.*

By assumption, $V_E$ is a vertex cover of $G$, and there exists some $\bar{v} \in V_E$ such that $V'_E = V_E/\{\bar{v}\}$ is also a vertex cover of $G$. Define $alloc' = \{B'_e\}_{e \in E} \cup \{B'_g\}$ as follows, with $V'_E$ and $V'_g$ corresponding to $alloc'$ as $V_E$ and $V_g$ do to $alloc$:

- For each edge $e \in E$, where $\bar{v} \notin e$, define $B'_e = B_e/\{\bar{v}', \bar{v}'', \bar{v}'''\}$. From the definition, $v_e(B'_e) = v_e(B_e)$.
- For each edge $e \in E$, where $\bar{v} \in e$, since $V'_E$ is a vertex cover, there exists some $\hat{v} \in (V'_E \cap e)$. Define $B'_e = \{f_{\hat{v}}(e)\}$. By construction, $v_e(B'_e) = 1 \geq \frac{v_e(B_e)}{2}$.
- Define $B'_g = B_g \cup \{\bar{v}', \bar{v}'', \bar{v}'''\}$. Therefore, $V'_g = V_g \cup \{\bar{v}\}$, and $v_g(B'_g) = \alpha \cdot v_g(B_g)$.

$alloc'$ partitions the items in $B$ to the agents in $A$, and is thus a valid allocation. Consequently, we have:

$$NSW(alloc') = \left( \prod_{e,\ v' \notin e} v_e(B'_e) \cdot \prod_{e,\ v' \in e} v_e(B'_e) \cdot v_g(B'_g) \right)^{\frac{1}{n}}$$

$$\geq \left( \prod_{e,\ v' \notin e} v_e(B_e) \cdot \prod_{e,\ v' \in e} \frac{v_e(B_e)}{2} \cdot \alpha \cdot v_g(B_g) \right)^{\frac{1}{n}}$$

$$= \left( \frac{8^N \cdot (c+\epsilon)^n}{2^3} \right)^{\frac{1}{n}} \cdot NSW(alloc) > c \cdot NSW(alloc)$$

contradicting the assumption that $alloc$ provides a $c$-approximation of the maximal Nash social welfare. ∎

**Notation 7.5.** Let $C^* \subseteq V$ be a minimum vertex cover of $G$ such that $C^* \in \arg\max_{C \in MVC} d_C$, where $MVC$ is the set of all minimum vertex covers of $G$.

**Lemma 7.6.** There exists an allocation $alloc$ such that $V_E = C^*$ and $NSW(alloc) = \left(2^{d_{C^*}} \cdot \alpha^{N-|C^*|}\right)^{\frac{1}{n}}$, which is the maximal Nash social welfare.

*Proof.*

Define $alloc = \{B_e\}_{e \in E} \cup \{B_g\}$ as follows:

- For each edge $e \in E$, define $B_e = \{f_v(e) | v \in C^* \cap e\}$. Since $C^*$ is a vertex cover, $C^* \cap e \neq \emptyset$, and thus, $v_e(B_e) \in \{1,2\}$.
- Define $B_g = B / \bigcup_{e \in E} B_e$. By construction, $v$ is covered by $B_g$ if and only if $v \notin C^*$. Therefore, $V_g = V/C^*$, and thus, $v_g(B_g) = \alpha^{N-|C^*|}$.

$alloc$ partitions the items in $B$ to the agents in $A$, and is thus a valid allocation. From the definition, there are exactly $d_{C^*}$ edges such that $|C^* \cap e| = 2$. By construction, there are exactly $d_{C^*}$ edge agents such that $v_e(B_e) = 2$. Altogether, we get that the Nash social welfare of $alloc$ is $\left(2^{d_{C^*}} \cdot \alpha^{N-|C^*|}\right)^{\frac{1}{n}}$.

Now, assume towards contradiction that there exists an allocation $alloc' = \{B'_e\}_{e \in E} \cup \{B'_g\}$, with $V'_E$ and $V'_g$ corresponding to $alloc'$ as $V_E$ and $V_g$ do to $alloc$, that achieves a higher Nash social welfare. By Lemmas 4.2 and 4.3, $NSW(alloc') = \left(2^k \cdot \alpha^{N-|V'_E|}\right)^{\frac{1}{n}}$, for some $0 \leq k \leq d_{V'_E}$, and $V'_E$ is a minimal vertex cover of $G$. Since $C^*$ is a minimum vertex cover of $G$, $|V'_E| \geq |C^*|$. If $|V'_E| > |C^*|$, by definition, $v_g(B'_g) = \frac{v_g(B_g)}{\alpha^{|V'_E|-|C^*|}}$, and then by assumption, $2^{k-d_{C^*}} > \alpha^{|V'_E|-|C^*|}$. Since the number of edges in $G$ is $\frac{3}{2}N$, we get that, $2^{k-d_{C^*}} \leq 2^{d_{V'_E}-d_{C^*}} < 8^N < \alpha \leq \alpha^{|V'_E|-|C^*|}$, which is a contradiction to the previous inequality. Hence, $|V'_E| = |C^*|$, and by definition, $d_{V'_E} > d_{C^*}$, and then $NSW(alloc) > NSW(alloc')$, which is a contradiction to the assumption. ∎

**Lemma 7.7.** Let $alloc = \{B_e\}_{e \in E} \cup \{B_g\}$ be an allocation such that $V_E$ is a minimal vertex cover of $G$ but not a minimum vertex cover of $G$. Then $alloc$ does not provide a $c$-approximation of the maximal Nash social welfare.

*Proof of Lemma 4.6.*

From Lemma 4.5, we know that the maximal Nash social welfare is $\left(2^{d_{C^*}} \cdot \alpha^{N-|C^*|}\right)^{\frac{1}{n}}$. We have

$$c \cdot NSW(alloc) \leq c \cdot \left(2^{d_{V_E}} \cdot \alpha^{N-|V_E|}\right)^{\frac{1}{n}} < \left(8^N \cdot c^n \cdot \alpha^{N-|V_E|}\right)^{\frac{1}{n}} < \left(\alpha^{N-|V_E|+1}\right)^{\frac{1}{n}}$$
$$\leq \left(2^{d_{C^*}} \cdot \alpha^{N-|C^*|}\right)^{\frac{1}{n}}$$

where the inequality derives from Lemma 4.2, the second inequality holds because the number of edges in $G$ is $\frac{3}{2}N$, and the last two from the definitions of $\alpha$ and $C^*$. Therefore, $alloc$ does not provide a $c$-approximation of the maximal Nash social welfare. ∎

By combining the statements from Lemmas 7.4 and 7.7, we conclude that if an allocation provides a $c$-approximation of the maximal Nash social welfare, $V_E$ must be a minimum vertex cover of $G$. From Lemma 7.6, we get that there exists an allocation providing a $c$-approximation of the maximal Nash social welfare where $V_E$ is a minimum vertex cover of $G$. The above two statements imply Theorem 6.1. ∎